\documentstyle[12pt]{article}
\input epsf
\newcommand{\eg}{{\em e.g.\ }}
\newcommand{\ie}{{\em i.e.\ }}
\newcommand{\vs}{{\em vs.\ }}
\begin{document}
\title {
\begin{flushright}
{\normalsize IIT-HEP-95/3\\
hep-ex/9508018\\
\vspace {-.15 in}
August 1995
}
\end{flushright}
\vskip 0.2in
\bf High-Impact Charm Physics at the Turn of the
Millennium\thanks{Presented at the
Workshop on the Tau/Charm Factory, Argonne National Laboratory,
June 21--23, 1995.}}
\author{ Daniel M. Kaplan
    \\ {\sl Illinois Institute of Technology, Chicago, IL 60616} 
	\\
         }  
\date{}
\maketitle

\begin{abstract}

I review the sensitivities achieved by and projected for fixed-target charm
experiments in {\em CP} violation, flavor-changing
neutral-current and lepton-number-violating decays, and mixing, and
I describe the Charm2000 experiment intended to run at Fermilab in the Year
$\approx$2000.
If approved, Charm2000 will in many of these areas
exceed the sensitivities projected for a Tau/Charm Factory, but the
Tau/Charm Factory retains certain qualitative advantages.

\end{abstract}

\section{Introduction}

\begin{figure}[htb]
\vspace{0.1in}
\hspace{-0.25 in}\centerline{\epsfysize=3 in\epsffile{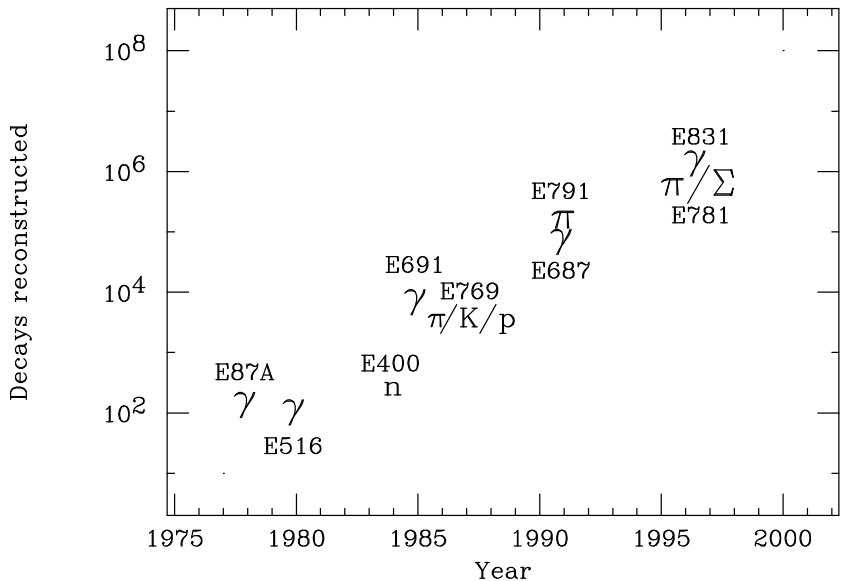}}
\caption{Yield of reconstructed charm \vs year of run
for those completed and approved Fermilab fixed-target charm experiments with
the highest statistics of their generation; symbols indicate type of beam
employed.\label{history}}
\end{figure}

A Tau/Charm Factory ($\tau c$F) may turn on early in the next millennium. At
that time one
can anticipate significant competition in charm physics from fixed-target
experiments, as well as from $e^+e^-$ colliders operating near $b{\bar b}$
threshold~\cite{Besson}. For many topics in charm physics (\eg lifetimes and
rare-decay searches), Fermilab fixed-target experiments now dominate the
field.
The progress of fixed-target charm experiments at Fermilab is sketched in
Fig.~\ref{history}, which shows roughly exponential
growth in sensitivity since the late 1970s.
While physics reach depends both on the number of
signal events reconstructed and on the amount of background under the peaks,
the former figure can still serve as a starting point for discussion. This
number is expected to reach $\sim$10$^6$ events during the next few years with
the runs of Fermilab E781 (SELEX) and E831 (FOCUS) and the advent of CLEO III.
In addition, a Letter of Intent has been submitted to CERN for an experiment
(CHEOPS) aiming to reconstruct $10^7$ charm~\cite{CHEOPS}, and one for a
$10^8$-charm experiment (Charm2000) at Fermilab in the Year $\approx\!2000$ is
in
progress~\cite{Kaplan2000,Kaplan95}. It is against this backdrop that the case
for a
Tau/Charm Factory must be evaluated.

\section{High-Impact Charm Physics}

``High-impact" denotes measurements which are particularly sensitive to new,
non-Standard-Model physics~\cite{Bigi89}.  The Standard Model (SM) contains two
key  mysteries: the origin of mass and the existence of multiple fermion
generations~\cite{Bigi94,Hewett} . We seek to answer the first in experiments
at  the LHC, exploring  the $\approx1\,$TeV mass scale of electroweak symmetry
breaking. The answer to  the second appears to lie at higher mass scales,
beyond what can be directly  accessed at the LHC. But these scales can be
probed in virtual loops in processes such as {\em CP}  violation, mixing, and
flavor-changing neutral or lepton-number-violating
currents~[7\,-\,9].

Such effects have been pursued with high sensitivity in the strange sector, and
in  the beauty sector they have become something of a holy grail, because of
large SM contributions to mixing and {\em CP} violation in the  decays  of
``down-type" quarks. These effects are enhanced for $s$ and $b$ quarks relative
to those for ``up-type"  quarks by the pattern of the Cabibbo-Kobayashi-Maskawa
(CKM) matrix and the  large mass of the top quark, whose contribution in loops
allows {\em CP}  violation by virtue of the CKM phase~\cite{Rosner}.  It is
precisely because these SM contributions  are {\em small} in the charm sector
that charm is a good place to look for  new-physics
contributions~[8,\,11\,-\,13]. Furthermore, charm is
the only
up-type quark for  which these studies are possible, since the top quark is
above $W+b$ threshold  and decays too quickly to form bound states.
The information available from charm studies is often complementary to that
from strangeness and beauty~\cite{Hewett,Yaouanc}.
Finally, as we
shall see, sensitivity  to new physics at interesting levels is anticipated in
upcoming charm  experiments: levels at which even the failure to observe an
effect imposes significant constraints on models.

Table~\ref{tab:sens} summarizes sensitivities in high-impact charm physics
currently achieved  and expected by the turn-on of the $\tau c$F, assuming
approval of the  Charm2000 project at Fermilab.  Table~\ref{tab:yields}
estimates yields of reconstructed events in various modes in Charm2000, some
directly and some by extrapolation from E791; since these yields vary
rapidly with vertex separation cuts, which are typically optimized differently
for each physics analysis, they are necessarily ill-defined at the factor-of-2
level.\footnote{Of course the statistical significance of signals, which
directly determines physics sensitivities, goes as the square root of yield
and is more stable with respect to cuts.} (To remind the reader of
this effect, I have indicated in
Table~\ref{tab:yields} the type of analysis for each E791 yield given.)
I next discuss each physics topic
in more detail,\footnote{The reach of Charm2000 in other physics areas
such as charm spectroscopy, tests of QCD, lifetimes, form factors, and
branching ratios will be discussed in a future publication.}
  following which I describe the salient aspects of the proposed
Charm2000  experiment.

\subsection{Direct {\em CP} violation}

The Standard Model predicts direct {\em CP} violation in
singly Cabibbo-suppressed decays (SCSD) of charm  at the $\sim$10$^{-3}$
level~[5,\,15\,-\,17],
arising from interference between tree-level and
penguin  diagrams for the decay of the charm quark. {\em CP} violation  in
Cabibbo-favored (CFD) or doubly Cabibbo-suppressed (DCSD) modes would however
be a clear signature for new physics~\cite{Burdman,Bigi94}. Asymmetries in all
three  categories could reach $\sim$10$^{-2}$ in such scenarios as
non-minimal
supersymmetry~\cite{Bigi94} and left-right-symmetric
models~\cite{Pakvasa,Yaouanc}. There are also expected SM asymmetries of
$\approx\!3.3\times10^{-3}$ ($=2\,Re(\epsilon_K)$) due to $K^0$ mixing in such
modes as $D^+\to K_S\pi^+$ and $K_S \ell\nu$~\cite{Xing}, which should be
observed in Charm2000 (Table~\ref{tab:sens}) or even in predecessor
experiments. While observation of  $K^0$-induced {\em CP} asymmetries might
teach us little new about physics, they will at least constitute a calibration
for the experimental systematics of  asymmetries at the $10^{-3}$ level.
However, Bigi has pointed out that a small new-physics contribution to the DCSD
rate could amplify these asymmetries to $\cal{O}$$(10^{-2})$~\cite{Bigi94}.

Experimental limits at the 10\% level have been set in SCSD modes; at present
the most sensitive
come from the photoproduction experiment Fermilab E687~\cite{Frabetti} and
from CLEO~\cite{Bartelt}. E687 has set
limits in $D^0\to K^+K^-$ and $D^+\to K^-K^+\pi^+$, $\overline {K^{*0}}K^+$,
and $\phi\pi^+$
as indicated in Table~\ref{tab:sens}.\footnote{To avoid such cumbersome
notations as
$D^0 (\overline {D^0})\to K^\mp\pi^\pm$, here and elsewhere in
this paper charge-conjugate states are generally implied even when not
stated.} CLEO has studied $D^0$ decays to the {\em CP} eigenstates $K^+K^-$,
$K_S\phi$,
and $K_S\pi^0$ as well as $K^\mp\pi^\pm$.

The signal for direct {\em CP} violation is an absolute rate difference between
decays of particle and antiparticle to charge-conjugate final states $f$ and
${\bar f}$:
\begin{equation}
A=\frac{\Gamma(D\to f)-\Gamma({\overline D}\to{\bar f})}
{\Gamma(D\to f)+\Gamma({\overline D}\to{\bar f})}\,.
\end{equation}
Since in photoproduction $D$ and ${\overline  D}$ are not produced equally, in
the E687 analysis the signal is normalized relative to the production asymmetry
observed in a CFD mode:
\begin {equation}
A=\frac{\eta(D\to f)-\eta({\overline  D}\to{\bar f})}
{\eta(D\to f)+\eta({\overline  D}\to{\bar f})}\,,
\end {equation}
where, for example,
\begin{equation}
\eta(D^0)=\frac{N(D^0\to K^+K^-)}{N(D^0\to K^-\pi^+)}\,,
\end{equation}
and for the $D^+$ modes the normalization mode is $D^+\to K^-\pi^+\pi^+$. (Thus
in the unlikely event that there is a {\em CP} asymmetry from new physics in
the CFD normalization mode which is equal to that in the corresponding
SCSD mode, the signal would be masked.) A further complication
is that to distinguish $D^0\to K^+K^-$ from $\overline {D^0}\to K^+K^-$,
$D^*$ tagging must be employed; of course, no tagging is needed for charged-$D$
decays. Typical E687 event yields are $\approx$10$^2$ in signal modes and
$\sim$10$^3$
in normalization modes.

Given the sensitivity achieved in E687, one can extrapolate to that
expected in Charm2000. E687 observed $4287\pm78$ ($4666\pm81$) events in the
normalization mode $D^+\to K^-\pi^+\pi^+$ ($D^-\to K^+\pi^-\pi^-$). As an
intermediate step in the extrapolation I use the event yield in E791, since
that hadroproduction experiment is more similar to Charm2000 than is E687.
Using relatively tight vertex cuts, E791 observed $37006\pm 204$ events in
$D^\pm\to K\pi\pi$~\cite{Aitala}, and Charm2000 should increase this number by
a factor $\approx$2000 (see Sec.~\ref{sec:yields}).
Thus relative to E687, the statistical
uncertainty on $A$ should be reduced by $\approx\!\sqrt{8000}$, implying
sensitivities in various modes of $10^{-3}$ at 90\% confidence. While  the
ratiometric nature of the measurement reduces sensitivity to systematic
biases, at the $10^{-3}$ level these will need to be studied carefully.

For DCSD modes, I extrapolate from E791's observation of $D^+\to
K^+\pi^+\pi^-$ at $4.2\sigma$ based on 40\% of their data
sample~\cite{Purohit-Weiner}. The statistical significance in Charm2000 should
be $\approx\!\sqrt{2000/0.4}$ better, implying few$\times10^{-3}$ sensitivity
for {\em CP} asymmetries. For $D^0\to K^+\pi^-$,
CLEO's observation~\cite{Cinabro} of $B(D^0\to K^+\pi^-)/B(D^0\to
K^-\pi^+)\approx0.8$\% suggests $\approx\!10^5$ $D^*$-tagged DCSD $K\pi$
events in
Charm2000, giving few$\times10^{-3}$ {\em CP} sensitivity. However, the need
for greater background suppression for DCSD compared to CFD events
is likely to reduce sensitivity. For example, preliminary E791 results show
a $\approx$2$\sigma$ signal in $D^0\to K^+\pi^-$~\cite{Purohit95},
implying $\sim$10$^{-2}$ sensitivity in Charm2000.
These extrapolations are conservative and ignore expected  improvements in
vertex resolution and particle identification. Detailed simulations  are
underway to assess these effects.

Sensitivities at a $\tau c$F have been estimated at a
few$\times10^{-3}$ in SCSD modes~\cite{Fry}, but
clear qualitative advantages make a $\tau c$F complementary to fixed-target
experiments~[25\,-\,27]. For example, the equal production of $D$
and ${\overline D}$ in $e^+e^-$ annihilation allows study of {\em CP} violation
at $<10^{-3}$ sensitivity in CFD modes,  a measurement which
in a fixed-target experiment can be carried out to greater statistical
precision (Table~\ref{tab:sens}) but depends on effects differing in size among
various CFD modes.
A $\tau c$F also has a clear advantage in modes with final-state photons.

SM predictions for direct {\em CP} violation are rather uncertain, since they
require assumptions for final-state phase shifts as well as CKM matrix
elements~\cite{Burdman,Bigi94}; the predictions given in Table~\ref{tab:sens}
are representative, but the theoretical uncertainties are probably larger than
indicated there~\cite{Buccella2}. However, given the order of magnitude
expected in charm decay, the Charm2000 experiment might make the first
observation of direct {\em CP} violation outside the strange sector, or indeed
the first observation anywhere if (as may well be the case~\cite{Paschos,Lu})
signals prove too small for detection in the next
round of $K^0$~[31\,-\,33] and hyperon~\cite{E871}
experiments.

\subsection{Flavor-Changing Neutral Currents}

Charm-changing neutral currents are forbidden at tree level in the Standard
Model due to the GIM mechanism~\cite{GIM}. They can proceed via loops at rates
which are predicted to be unobservably small, \eg for $D^0\to\mu^+\mu^-$
(which suffers also from helicity suppression in the SM) the predicted
branching ratio is $\sim$10$^{-19}$~\cite{Gorn,Pakvasa,Hewett}, and for
$D^+\to\pi^+\mu^+\mu^-$ it is $\sim$10$^{-10}$~\cite{Babu,Hewett}.
Long-distance  effects increase these predictions by some orders of magnitude,
but they  remain of order $10^{-15}$ to
$10^{-8}$~\cite{Pakvasa,Schwartz,Burdmanetal}. Various extensions of the
SM~\cite{Babu,Lepto} predict effects substantially
larger than this, for example in models with a fourth generation, both
$B(D^+\to\pi^+\mu^+\mu^-)$ and $B(D^0\to\mu^+\mu^-)$ can be as large as
$10^{-9}$~\cite{Babu}. Experimental sensitivities are now in the range
$\sim$10$^{-4}$ to $10^{-5}$~[21,\,40\,-\,43]
and are expected to reach $\sim$10$^{-5}$ to $10^{-6}$ in E831~\cite{Cumalat}.

Limits on FCNC charm decays have recently improved considerably, with new
results from Fermilab E653 and E791 and WA92 at CERN.
E653~\cite{E653} studied charm
decays to  hadrons plus muon pairs in a variety of modes, E791~\cite{Aitala}
studied  charged-$D$ decays
to $\pi\mu^+\mu^-$ and $\pi e^+ e^-$, and WA92~\cite{WA92} searched for
$D^0\to\mu^+\mu^-$. Typically a normalization mode is used to determine the
sensitivity, reducing systematic uncertainty. Thus E791
normalized to $K^-\pi^+\pi^+$ and WA92 to $K^\mp\pi^\pm$, eliminating
normalization uncertainty due to the $D$ production cross section. (Older
limits~\cite{Louis,Mishra} on $D^0\to\mu^+\mu^-$ used $J/\psi\to\mu^+\mu^-$ for
normalization, reducing uncertainty due to muon identification and triggering
efficiency.)

One can extrapolate from recent results to estimate sensitivities in
Charm2000. While Charm2000 aims at a single-event branching-ratio sensitivity
of $\approx\!10^{-9}$, FCNC limits are typically background-limited, so
sensitivites can be expected to improve as the square root of the number
of events reconstructed. In some cases, however, more dramatic improvement may
result from
improved lepton identification. For $D^+\to\pi^+\mu^+\mu^-$, scaling the
E791 sensitivity by a factor $\sqrt{2000}$ as above gives
$\approx$few$\times10^{-7}$ 90\%-confidence sensitivity in Charm2000. This
estimate may
be conservative, since the simple muon detection scheme employed by E791 (one
layer of scintillation counters following 2.5\,m of steel equivalent) resulted
in a (momentum-dependent)  $\pi$-$\mu$ misidentification probability ranging
from 4.5 to 20\%~\cite{Aitala},
and it should be possible to reduce this to $\approx$1\% in
Charm2000. With modern calorimetry for electron identification one expects to
do almost  as well for $\pi e e$ as for $\pi\mu\mu$. For $D^0\to\mu^+\mu^-$ and
$e^+e^-$,  extrapolating from WA92 implies sensitivity of $10^{-7}$ per mode.

Radiative charm decays present the opportunity to test models
of nonperturbative long-distance effects, since short-distance (penguin)
contributions are estimated to be negligible even in extensions of the SM such
as models with a fourth generation~\cite{Burdmanetal}.
Long-distance effects give
branching ratios of order $10^{-5} - 10^{-6}$,
whereas current experimental limits are $\sim$10$^{-4}$
(see Table~\ref{tab:sens}). It is important to test
these calculations in the charm sector, where the predicted effects are large
and not ``contaminated" by short-distance physics,  since small but
non-negligible long-distance  corrections  are predicted in the $b$ sector,
where \eg one would like to extract the CKM element $V_{td}$ from
$B(B\to\rho\gamma)$~\cite{Bigi2000,Hewett,Ali}.
In addition there may be a window for new
physics, since \eg non-minimal supersymmetry might make a substantial
contribution to $D\to\rho\gamma$, and this may be distinguishable from a
long-distance SM effect since the latter is Cabibbo-suppressed with respect to
$D^0\to K^*\gamma$ in the SM but not in SUSY~\cite{Bigietal,Bigi2000}.
Observation of such modes as $D^0\to \rho^0\gamma$ and
$D^0\to\phi\gamma$ may be within reach at a $\tau c$F or $B$ factory. It is not
clear how well
fixed-target experiments can do on these modes, since they
must cope with large combinatoric photon backgrounds from $\pi^0$ decay.

\subsection{Lepton-Number-Violating Decays}

There are two lepton-number-violating effects which can be sought:
decays violating conservation of lepton number (LNV) and decays violating
conservation of lepton-family number (LFNV). LFNV decays (such as
$D^0\to\mu^\pm e^\mp$) are expected in theories with leptoquarks~\cite{Lepto},
heavy neutrinos~\cite{Hewett}, extended technicolor~\cite{Techni}, etc.;
LNV decays (such as $D^+\to K^- e^+ e^+$) can arise in GUTs and have been
postulated to play a role in the development of the baryon asymmetry of the
Universe~\cite{cosmic-baryon}. Since no fundamental principle forbids
either type of decay, it is of interest to search for
them as sensitively as possible.

Although much smaller decay widths can be probed in $K$ decays, there are
simple theoretical arguments why LFNV charm decays are nevertheless
worth seeking. If such currents arise through Higgs
exchange, whose couplings are proportional to mass, they will couple more
strongly to charm than to strangeness~\cite{Bigi87}.
Furthermore, LFNV currents
may couple to up-type quarks more strongly than to
down-type~\cite{Lepto,Hadeed}.

As shown in Table~\ref{tab:sens}, the best existing limits come in most cases
from the $e^+e^-$ experiments Mark II, ARGUS, and CLEO (although the
hadroproduction experiment Fermilab E653 dominates in modes with same-sign
dimuons) and are typically at the $10^{-3}-10^{-4}$ level~\cite{Sheldon,E653}.
E831 expects to lower these limits to $\sim$10$^{-6}$~\cite{Cumalat},
and Charm2000 should reach $\sim$10$^{-7}$.

\subsection{Mixing and Indirect {\em CP} Violation}

$ D^0\overline {D^0} $ mixing may be one of the more promising places to
look for low-energy manifestations of physics beyond the Standard Model.
For small mixing, the mixing rate is given to good approximation
by~\cite{Burdman}
\begin{equation}
r_{\rm mix}\approx \frac{1}{2}\bigg [\bigg (\frac{\Delta M_D}{\Gamma_D}
\bigg )^2 + \bigg (\frac{\Delta\Gamma_D}{2\Gamma_D}\bigg )^2\bigg ]\,.
\end{equation}
In the SM the $\Delta M$ and $\Delta\Gamma$
contributions are expected~\cite{Burdman} to be of the same order of
magnitude and are estimated~\cite{Burdman,mixing} to give
$r_{\rm mix}<10^{-8}$; any observation at a
substantially higher level will be clear evidence of new
physics.\footnote{Earlier estimates~\cite{Wolfenstein_mixing}
that long-distance effects can give $|\Delta M_D/\Gamma_D |\sim 10^{-2}$
are claimed to have been disproved~\cite{Burdman,Hall-Weinberg}, but there
 remain skeptics~\cite{Bigi94,Wolfenstein}.}
Many nonstandard models predict much larger effects.
An interesting example  is the multiple-Higgs-doublet model
lately expounded by Hall and Weinberg~\cite{Hall-Weinberg}, in  which
$|\Delta M_D|$ can be as large as $10^{-4}$\,eV, approaching the current
experimental limit. In this model all {\em CP} violation arises from
Higgs exchange and is intrinsically of order 10$^{-3}$, too small to be
observed in the beauty sector and (except through mixing) in the kaon sector,
but possibly observable in charm -- another example of the importance of
exploring rare phenomena in {\em
all} quark sectors. The large mixing contribution arises from flavor-changing
neutral-Higgs exchange (FCNE)~\cite{Wu},
which can be constrained to satisfy the GIM
mechanism for $K^0$ decay by assuming small phase factors ($\sim$10$^{-3}$).
(This is in distinction to the original ``Weinberg model" of {\em CP}
violation~\cite{Weinberg}, in which FCNE was suppressed by assuming a discrete
symmetry such that one Higgs gave mass to up-type quarks and another to
down-type.) Multiple-Higgs models are one of the simplest extensions of
the SM~\cite{Pakvasa,Winstein-Wolfenstein,Wu}, and many other authors have also
considered multiple-Higgs effects in charm mixing~[49,\,52,\,59\,-\,61].
Large mixing in charm can also arise in theories with
supersymmetry~\cite{Bigietal,Datta,Nir}, technicolor~\cite{Techni},
leptoquarks~\cite{Lepto}, left-right symmetry~\cite{seesawLR},
or a fourth generation~\cite{Pakvasa,Babu}.

The experimental situation regarding $D^0\overline {D^0}$ mixing is complicated
by the presence of DCSD. Since both effects can lead to the same final states,
one needs to distinguish them using time-resolved measurements~\cite{Bigi87}.
In the notation of Refs.~\cite{Blaylock} and \cite{Browder}, the time
dependence for wrong-sign decay is given by
\begin{eqnarray}
\label{eq:mixing}
\Gamma(D^0(t)\to K^+\pi^-)=\frac{e^{-\Gamma t}}{4}
|B|^2|\frac{q}{p}|^2 \{4 |\lambda|^2
+ (\Delta M^2 +\frac{\Delta\Gamma^2}{4})t^2 + \nonumber \\
2Re(\lambda)\Delta\Gamma t + 4 I\!m(\lambda)\Delta Mt\}\,,
\end{eqnarray}
and there is a similar expression for $\overline{D^0}\to K^-\pi^+$
in which $\lambda$ is replaced by ${\bar \lambda}$.
In Eq.~\ref{eq:mixing} the first term on the right-hand side is the DCSD
contribution, which peaks at $t=0$;
the second is the mixing contribution, which peaks at 2 $D^0$
lifetimes because of  the factor $t^2$; and the third and fourth
terms reflect interference between mixing and DCSD and peak at 1 lifetime due
to the factor $t$.
$\lambda$ and $\bar \lambda$ can acquire nonzero phases through indirect {\em
CP} violation or through final-state interactions~\cite{Browder,Wolfenstein}.
While for sufficiently small $|\Delta M/\Gamma|$ experimental sensitivity to
mixing is enhanced if there is interference~\cite{Liu}, at present levels of
sensitivity allowing an arbitrary interference phase when fitting decay-time
distributions reduces the stringency of the resulting
limit~\cite{E691,Purohit95}.

The most sensitive limit on $D^0\overline {D^0}$
mixing (quoted in Table~\ref{tab:sens} and in the
{\it Review of Particle  Properties}~\cite{PDG}) comes from the Fermilab
photoproduction experiment E691~\cite{E691}.
The E691 analysis considered two modes, $D^0\to K^\mp\pi^\pm$ and $K^\mp\pi^\pm
\pi^+\pi^-$, and five possible values of the interference phase $\phi$ covering
the range $-1\leq \cos{\phi}\leq 1$. The limits in each mode were stable over
most of the $\phi$  range, but worsened for $\cos{\phi}= -1$ by a factor 1.8
(3.3) for $K\pi$ ($K 3\pi$). The final result was derived by combining the two
modes neglecting interference.

Recently several authors have
critiqued the E691 mixing analysis. Liu~\cite{Liu} has questioned the validity
of the combined limit, suggesting that even if interference is negligible for
one mode, it is less likely to be negligible for both. Blaylock,
Seiden, and Nir~\cite{Blaylock} and Wolfenstein~\cite{Wolfenstein}
suggest that whereas the E691 fit neglected the term in
Eq.~\ref{eq:mixing} proportional to $\Delta M$ but kept the term in
$\Delta \Gamma$, the reverse should have been done.
Browder and Pakvasa~\cite{Browder} have reconsidered the E691
analysis taking into account the role of final-state interactions;
they conclude that even maximal destructive
interference degrades the no-interference E691 limit only at the 10\% level.
However, the understanding of
final-state phases is entirely phenominological, and more
work and data are required to assess its reliability.
Nevertheless it appears that the E691 limit is not ``wrong" by much
if at all, and interference does not appear to play a large role at present
sensitivity.

While there is as yet no published mixing limit from E791, the preliminary
indication is sensitivity to $r$
at the $\approx$10$^{-3}$ level if interference is
neglected, ranging to perhaps a few times this if interference is
allowed~\cite{Purohit95}.
A simple extrapolation by $\sqrt{2000}$ suggests sensitivity of
$\approx$2$\times10^{-5}$ in Charm2000 neglecting interference, which with
improvements in particle identification and resolution for the tagging pion
might approach $10^{-5}$. However,
since the interference term is linear in $\Delta M_D$ while the mixing term is
quadratic, the ratio of the interference and mixing contributions
goes as $1/\Delta M_D$. Thus as experimental sensitivity
improves and smaller and smaller values of $|\Delta M_D|$ are probed,
interference becomes relatively more important.
One therefore cannot extrapolate simply from the E691 or E791 sensitivity
to that expected in Charm2000.

A first attempt to assess the impact of interference on mixing sensitivity in
Charm2000 has been carried out by generating ten Monte Carlo samples of DCSD
$D^0\to K^+\pi^-$ events and fitting them allowing for interference or not.
I conservatively assume $10^4$ events observed after vertex cuts and fit each
decay-time histogram only for $t>0.88\,$ps (2 $D^0$ lifetimes) as in the E691
analysis, following the
prescription of Browder and Pakvasa~\cite{Browder} for the time dependence in
the case of no {\em CP}
violation (their Eq.~4). Within their suggested range of final-state phase
(5$^\circ$ to 13$^\circ$), the interference term improves sensitivity slightly,
and $10^{-5}$ sensitivity is obtained.
Since the interference contribution peaks at 1 lifetime it would be desirable
to include shorter decay times in the fit, however more simulation studies are
required to evaluate signal cleanliness in that region.

Semileptonic decays offer a way to study mixing free from the effects of DCSD.
So far the only published limit on
charm mixing from semileptonic decays (Table~\ref{tab:sens})
is from the Fermilab dimuon hadroproduction
experiment E615~\cite{Louis}, in which only the muons were detected and no
vertex information was available.
A preliminary result from E791 using $D^*$-tagged $D^0\to K e\nu$ events
indicates sensitivity at the $\approx$0.5\% level~\cite{Tripathi}.
Extrapolation by $\sqrt{2000}$ suggests $10^{-4}$ sensitivity in
Charm2000, but use of muonic decays as well, plus improvements in lepton
identification and resolution for the tagging pion,
may give significantly better sensitivity.
At the Charm2000 Workshop, Morrison suggested $10^{-5}$ sensitivity may be
possible~\cite{Morrison}.

Liu has stressed the importance of setting limits on $\Delta\Gamma$ as well as
on $\Delta M$. Although typical extensions of the SM which predict large
$|\Delta M|$ also predict $|\Delta
M|\gg |\Delta\Gamma|$~\cite{Blaylock,Browder},
from an experimentalist's viewpoint both should be measured if possible.
$\Delta\Gamma$ can be studied quite straightforwardly by comparing the lifetime
measured
for {\em CP}-even modes such as $K^+ K^-,\pi^+\pi^-$ with that for {\em
CP}-odd modes or (more simply) with modes of mixed {\em CP} such as $K^-\pi^+$.
No such result has yet been published, so it is difficult to extrapolate
realistically to Charm2000 sensitivity.
Liu~\cite{Liu} has estimated Charm2000 sensitivity (in an idealized case)
at $\sim 10^{-5}-10^{-6}$ in $(\Delta\Gamma/2\Gamma)^2$ (\ie the contribution
to $r$ due to $\Delta\Gamma$).

The $\tau c$F can make a unique contribution to the study of mixing. DCSD
are forbidden in decays such as $\psi^{\prime\prime}\to D^0 \overline{D^0}
\to (K^-\pi^+)(K^-\pi^+)$ due to the  $C=-1$ initial state and
the Bose symmetry of the final state~\cite{Bigi-Sanda,Yamamoto,Gladding},
allowing
direct time-integrated observation of mixing in hadronic final states;
sensitivity has been estimated at $\sim$10$^{-4}$ per year of
running~\cite{Gladding}.

\subsubsection{Indirect {\em CP} violation}
\label{indirect}

Since in the SM $D^0\overline {D^0}$ mixing is negligible,
any indirect {\em CP}-violating asymmetries are
expected to  be less than $10^{-4}$~\cite{Bigi94}.
However, possible mixing signals at the $\approx$1\% level
have been reported~\cite{Cinabro,MkII-mixing}.
While given the E691 limit these probably represent enhanced DCSD signals,
if a significant portion of this rate is in fact mixing then new physics must
be responsible~\cite{Burdman,Wolfenstein}.
Then indirect {\em CP} violation at the $_\sim$\llap{$^<$}1\%
level is possible~\cite{Bigi-Sanda,Bigi2000,Bigi94,Wolfenstein}.
Some authors have suggested that the {\em CP}-violating signal, which
arises from the interference term of Eq.~\ref{eq:mixing}, may be
more easily detectable than the mixing
itself~[56,\,64\,-\,66].
In particular, Browder and
Pakvasa~\cite{Browder} point out that in the difference
$\Gamma(D^0\to K^+\pi^-)-\Gamma(\overline{D^0} \to K^-\pi^+)$, the DCSD
and mixing components cancel, leaving only the fourth term of
Eq.~\ref{eq:mixing}. Thus if indirect {\em CP} violation is
appreciable this is a particularly clear way to isolate the
interference term.

\section{A Next-Generation Charm Spectrometer}

A Letter of Intent is in progress for an experiment which can achieve the
$10^8$-reconstructed-charm sensitivity mentioned above. As we will see, the
most demanding requirement is on the trigger. In particular, an on-line
secondary-vertex trigger is needed if adequate trigger rejection is to be
achieved without sacrificing sensitivity in  hadronic decay modes.
(More detailed discussions may be found in \cite{Kaplan2000} and
\cite{Kaplan95}.)

\subsection {Beam and target}

To achieve  the sensitivity discussed here in a fixed-target run of
$\approx$10$^5$ beam spills requires a primary
proton beam~\cite{architecture}. Assuming 800-GeV beam energy the
charmed-particle production rate is
$7\times10^{-3}$/interaction if a high-$A$ target (\eg Au) is
used, or $3\times10^{-3}$ if diamond is used~\cite{world-average}.

A target which is short compared to typical charm decay lengths is crucial for
optimizing background suppression, both off-line and at trigger level. While
multiple thin targets could be employed (as in E791 and E831),
a single  target facilitates fast vertex triggering. A $\approx$1\,mm W, Pt,
or Au
target is one possibility, representing $\approx$1\% of an interaction length
and on average $\approx$15\% of a radiation length for outgoing secondaries. A
low-$Z$
material  such as diamond may be favored to minimize scattering of
low-momentum pions from $D^*$ decay~\cite{architecture};
then a $\approx$2\,mm  target is suitable,
representing $\approx$1\% of an interaction length and $\approx$1\% of a
radiation length. Given the mean Lorentz boost $\gamma\approx
35$,  a 1--2\,mm target is short enough that a substantial fraction even of
charmed baryons will decay outside it.

For triggering purposes (see Sec.~\ref{sec:trig}) and to optimize resolution in
decay distance, it is desirable to minimize the rate of
occurrence of multiple simultaneous interactions.
We therefore assume a 5\,MHz interaction rate, which given the Tevatron's
53\,MHz bunch rate and the typical 50\% spill duty factor implies a
$\approx$20\% fraction of events with multiple interactions. The needed
0.5--1\,GHz of primary proton beam is easily attainable.
As shown in Sec.~\ref{sec:yields}, this
yields  $_\sim$\llap{$^>$}\,$10^8$ reconstructed charm per
few\,$\times\,10^6$\,s of
beam ($\approx$10$^5$ spills $\times~20\,$s/spill).

\subsection {Spectrometer}

\begin{figure}[htb]
\centerline{\epsfysize = 1.89 in \epsffile {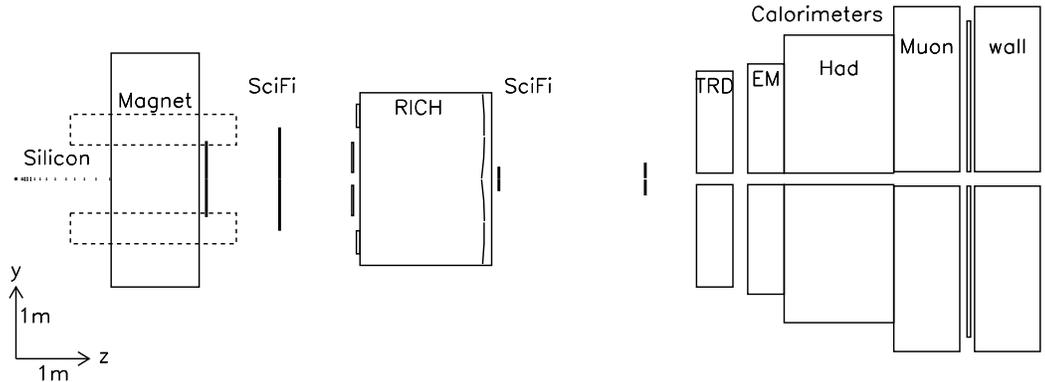}}
\caption [Spectrometer layout (bend view).]%
{Spectrometer layout (bend view).}
\label{app865}
\end{figure}

\begin{figure}[htb]
\vspace{0.1in}
\hspace{-.1in}\centerline{\epsfysize = 1.8 in \epsffile {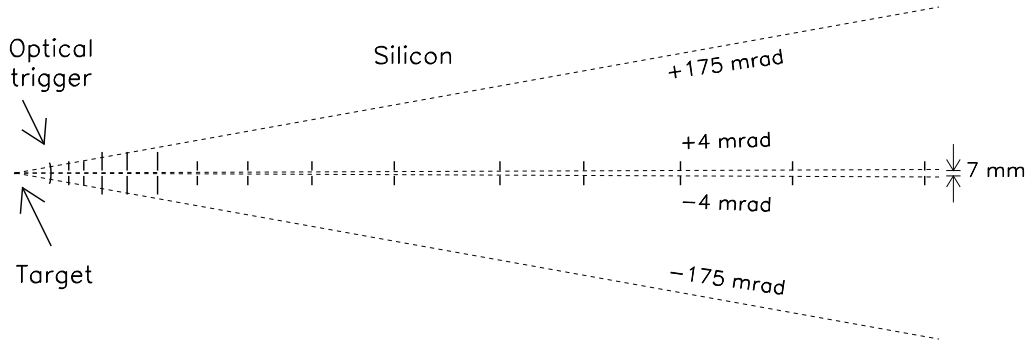}}
\caption [Detail of vertex region (showing optional optical impact-parameter
trigger).]%
{Detail of vertex region (showing optional optical impact-parameter
trigger).}
\label{detail865}
\end{figure}

We assume a highly rate-capable large-acceptance open-geometry spectrometer.
A significant design challenge is posed by radiation damage to the vertex
detectors. To configure detectors which can survive at the desired
sensitivity, we choose suitable maximum and (in one view) minimum angles for
the instrumented aperture, arranging the detectors along the beam axis with a
small gap through which pass the uninteracted beam and secondaries below the
minimum angle (Figs.~\ref{app865},~\ref{detail865}).\footnote{An
alternative approach with no gap may also be workable if the beam is spread
over sufficient area to satisfy rate and radiation-damage limits, however the
approach described here
probably allows smaller vertex detectors and is ``cleaner" in that the beam
passes through a minimum of material.} Thus the rate is spread approximately
equally over several detector planes, with
large-angle secondaries measured close to the target and small-angle
secondaries farther downstream. Along the beam axis the spacing of detectors
increases geometrically, making the lever arm for vertex
reconstruction independent of production angle. Since small-angle secondaries
tend to have high momentum, the multiple-scattering contribution to vertex
resolution is also approximately independent of production angle. We have
chosen an instrumented angular range $|\theta_x| \leq 200\,$mr, $4\leq
\theta_y \leq 175\,$mr, corresponding to the center-of-mass rapidity range
$|y_{\rm CM}|\,_\sim$\llap{$^<$}$\,1.9$ and containing over 90\% of produced
secondaries.

Assuming $n$ charged particles per unit pseudorapidity, the rate per unit
detector area at transverse distance $r$ from the beam is given by $n/2\pi
r^2$.
Since in 800\,GeV  proton-nucleus collisions
$n\approx4$ for high-$A$ targets~\cite{HERA-B} (less for C), in a run of
$n_{\rm int}$ interactions, a detector which can withstand a maximum fluence
of $R_{\rm max}$ particles/cm$^2$ has a ``minimum survivable" inner detector
radius
\begin{equation}
r_{\rm min}= {\left ( {n\over2\pi}{n_{\rm int}\over{R_{\rm max}}}\right )}
^{1\over2}\,.
\end{equation}
A typical run will yield fewer than
$2\times10^{13}$ interactions. If we assume currently-available
silicon detectors ($R_{\rm max} \approx 10^{14}/$cm$^2$), we obtain
conservatively
$r_{\rm min}=3.5$\,mm. An order-of-magnitude improvement in radiation hardness
would reduce $r_{\rm min}$ to $\approx$1\,mm, which is close to the minimum
half-gap through which the beam could be reliably steered. In
Fig.~\ref{detail865} we have conservatively indicated
3.5\,mm as the half-gap. Vertex resolution tends to improve as the half-gap is
reduced, so the use of radiation-hard detectors (either diamond
detectors~\cite{Tesarek} or improved silicon detectors) is highly desirable;
such detectors are likely to be available by $\approx$Year 2000. The desired
angular range can be covered with sufficient redundancy for pattern
recognition using 14 double-sided vertex detectors above and 14 below the beam
as
shown in Fig.~\ref{detail865}. These might be radiation-hard silicon-strip or
-pixel or diamond-strip or -pixel detectors.

Downstream of the analyzing magnet we assume scintillating-fiber tracking
using 3HF/PTP fibers with VLPC readout~\cite{Atac3} as in the D0~\cite{Ruchti}
and CDF upgrades.
The minimum half-gap for the fiber planes is determined by occupancy, which in
the uniform-pseudorapidity approximation used above (and neglecting magnetic
bending) is given by
\begin{equation}
{n\over\pi}{dy\over y}\arctan{x_{\rm max}\over y}
\end{equation}
for a detector element of height $dy$ a distance $y$ from the beam which
covers $-x_{\rm max}<x<x_{\rm max}$. For 800\,$\mu$m fiber
diameter, this implies $\approx$16\% occupancy at $y=1\,$cm,
$\approx$8\% at 2\,cm, and $\approx$4\% at
4\,cm. A full trackfinding
simulation will be required to assess the maximum acceptable occupancy,
but this suggests $\approx$1\,cm as the minimum acceptable half-gap in the
scintillating-fiber planes. The fibers near the gap could be split at
$x=0$ and read out at both ends, halving their occupancies. Since shorter
fibers have less attenuation, a smaller diameter could be used near the
gap, reducing occupancy still further.
Since the fibers are more radiation-hard than silicon
detectors and the fiber-plane beam gap
is larger than that of the vertex detectors,  radiation damage of the fibers
will not
be a problem.

The spectrometer sketched here accepts $_\sim$\llap{$^>$}\,50\% of two-prong
$D^0$ decays and $\approx$50\% of three-prong decays, comparable to E687 and
E791 acceptances. Assuming a 0.5\,GeV analyzing-magnet $p_t$ kick, the $D$
mass resolution ($\approx$5\,MeV rms) is a factor $\approx$2 better than that
of E687 or E791.
With vertex detectors of 25\,$\mu$m pitch read out digitally (\ie no
pulse-height information), vertex resolution is comparable to that of existing
spectrometers; we are exploring the possible improvement from reduction of the
half-gap to 1 mm and use of analog readout via flash ADCs as in E831.
(Since the mass resolution is dominated by scattering,
minimization of material is crucial,
for example use of helium bags and avoidance of threshold Cherenkov counters
employing heavy gas mixtures.)

\subsection{Trigger}
\label{sec:trig}

While previous Fermilab charm hadroproduction experiments E769 and  E791
recorded and analyzed large charm samples using very loose
triggers which accepted most inelastic interactions, this
approach is unlikely to extrapolate successfully by three orders of magnitude!
(Consider that E791 recorded $2\times10^{10}$ events -- 50 terabytes of
data -- on 24,000 8\,mm tapes.) Thus our sensitivity goal requires a highly
selective trigger. However, we wish to trigger on charm-event characteristics
which bias the physics as little as possible.
We therefore assume a first-level trigger requiring
calorimetric $E_t$ (as in E769 and E791)
OR'ed with high-$p_t$-lepton and lepton-pair triggers.
At second level, secondary-vertex requirements are imposed
on the $E_t$-triggered events to achieve a rate
($\sim$100\,kHz) which is practical to record.

\subsubsection{$E_t$ trigger}

Based on experience in E791, and using PYTHIA to simulate the effect of
pile-up in the calorimeter\footnote{Given 20\%
probability for $>$1 interaction, pile-up degrades the rejection by a factor
$\approx$2.}~\cite{Karchin}, we
expect a $\approx$10\,GeV $E_t$
threshold to give a minimum-bias rejection factor of 5 with $\approx$50\%
charm efficiency. (These are rough estimates based on a relatively crude
calorimeter, and an optimized calorimeter may provide better
rejection.) Such an $E_t$ trigger yields a 1\,MHz input rate to the next
level; the leptonic trigger rates should be negligible by comparison.

\subsubsection{Secondary-vertex trigger}

An additional factor $\approx$10 in trigger rejection is
desirable, and can be achieved by requiring evidence of secondary vertices.
This might be accomplished using a hardware trigger processor, which would
need to be an order of magnitude faster than existing vertex
processors~\cite{Processors} to accept events at 1 MHz; fast readout and
buffering  of event information would also be required. Trackfinding
secondary-vertex triggers benefit from the use of focused beam and a single
thin target, which allow simplification of the algorithm since the primary
vertex location is known {\it a priori}.

Christian~\cite{Christian} has suggested a simple trigger-processor
algorithm based on this idea. A PYTHIA-based simulation of this
algorithm for the vertex-detector configuration of
Fig.~\ref{detail865} shows good performance~\cite{Kaplan95}.
Assuming negligible spread in $y$
of primary-interaction vertices,\footnote{achievable \eg by use a target of
$100\,\mu$m height.}
 requiring at least one track to miss the
primary vertex by at least 200$\,\mu$m in $y$ rejects 95\% of minimum-bias
events while retaining 67\% of all charm events.
The simulation also tested the effects of making a preliminary pass through
the data eliminating hits which lie on straight lines pointing to the primary
vertex: rejection and efficiency were hardly affected. Since as the number of
hits per detector plane ($n$) increases, the
time to eliminate hits is linear in $n$, while the time to find tracks of
finite impact parameter goes as $n^2$ (due to the required loops over hits in
two seed
planes), such a hit-elimination pass can reduce processing time
substantially~\cite{Christian}.

As alternatives to iterative trackfinding at a 1\,MHz event rate,
three other approaches also appear worth pursuing.  The first is a
secondary-vertex trigger implemented using fast parallel logic, \eg PALs,
neural networks, or pre-downloaded fast RAMs, to look
quickly for patterns in the vertex detectors
corresponding to tracks originating downstream of the target. The others are
fast secondary-vertex trigger devices originally proposed for beauty: the
optical impact-parameter~\cite{optrig} and Cherenkov
multiplicity-jump~\cite{mul-jump} triggers;  while results from prototype tests
so far suggest  lower than desired charm efficiency, these might with further
development provide  sufficient resolution to trigger efficiently on charm. For
example, one  simulation of an optical impact-parameter trigger~\cite{P865}
indicated 40\%  charm efficiency for a factor 5 minimum-bias rejection, which
is good enough  to be usable in Charm2000.  In a very different regime of decay
length and impact parameter,  an optical trigger is in development for the
hyperon {\em CP}-violation  experiment Fermilab E871~\cite{E871optrig};
experience gained from this effort should allow prediction of charm performance
with good confidence. A charm multiplicity-jump trigger is under development
for CHEOPS~\cite{CHEOPS,Kwan}.

\section{Yield}
\label{sec:yields}

The charm yield is straightforwardly estimated. Assuming a Au target and a
typical fixed-target run of $3\times10^6$ live beam seconds, $10^{11}$ charmed
particles are produced. The reconstructed-event yields in
representative modes are estimated
in Table~\ref{tab:yields} assuming (for the sake of illustration)
that the optical trigger described in \cite{P865} is used for all-hadronic
modes (but not for leptonic modes, for which the first-level
trigger rate should be sufficiently low to be
recorded directly) and performs as estimated
above. Although due to off-line selection cuts not yet simulated,
realistic yields could be a factor $\approx2-3$ below those indicated,
the total reconstructed sample is
well in excess of $10^8$ events. Given the factor $\approx$2
mass-resolution improvement
compared to E791, one can infer a factor $\approx$50 improvement in
statistical significance in typical decay modes.

\section{Conclusions}

A fixed-target hadroproduction experiment (Charm2000)
capable of reconstructing in
excess of $10^8$ charm events is feasible using detector, trigger, and data
acquisition technologies which exist or are under development.
A typical factor $\approx50$ in statistical
significance of signals may be expected compared to E791.
We expect the spectrometer sketched here to cost substantially less than
HERA-$B$ (whose cost was estimated at 33M DM in
1994~\cite{HERA-B}).\footnote{While HERA-$B$ is potentially competitive with
Charm2000 as a charm experiment, it lacks the capabilities to trigger
efficiently on charm and to acquire the needed large data sample, and it
probably has significantly poorer vertex resolution as well.}
Should such an experiment be carried out it will likely exceed the sensitivity
of a
$\tau c$F in the high-impact areas of charm {\em CP} violation, mixing, and
flavor-changing neutral and lepton-number-violating  currents.
This conclusion might be questioned in light of recent scheduling experience
at Fermilab. However, the typical $\approx$3-year interval between Fermilab
fixed-target runs is offset by the need to divide $\tau c$F running time among
various physics topics requiring differing beam energies.\footnote{The
frequency of Fermilab fixed-target runs might also increase once Main Injector
construction is completed.}
Even without Charm2000, the CHEOPS experiment may come within an order of
magnitude of Charm2000 sensitivity and rival that achievable in a $\tau c$F.
Neverthelesss, the $\tau c$F complements charm hadroproduction experiments by
its capability to make various unique measurements, not to mention its
capabilities in $\tau$ physics~\cite{Huang}.
Ideally, both projects will go forward.

\section*{Acknowledgements}

I thank J. A. Appel, I. I. Bigi, C. N. Brown, G. Burdman, D. C. Christian, J.
L. Hewett,
S. Kwan, T. Liu, S. Pakvasa, and M. D. Sokoloff for useful discussions, and J.
Repond for the invitation to participate in this Workshop.
The organizers deserve particular thanks for a memorable workshop dinner.

{\footnotesize
\begin{table}
\caption{Sensitivity to high-impact charm physics. \label{tab:sens}}
\begin{center}
\begin{tabular}{|l|l|l|l|l|}
\hline
& & Charm2000 & SM \\
\raisebox{1.5ex}[0pt]{Topic} & \raisebox{1.5ex}[0pt]{Limit$^*$} & Reach$^*$
& prediction \\
\hline\hline
Direct {\em CP} Viol.\
& & & \\ \hline
{}~$D^0\to K^- \pi^+$ & -0.009$<$$A$$<$0.027~\cite{Bartelt} &  & $\approx0$
(CFD)
\\
{}~$D^0\to K^- \pi^+\pi^+\pi^-$ & & few$\times10^{-4}$ & $\approx0$ (CFD)
\\
{}~$D^0\to K^+ \pi^-$ &
& $10^{-3}-10^{-2}$ & $\approx0$ (DCSD) \\
{}~$D^+\to K^+ \pi^+ \pi^-$ &
& few$\,\times10^{-3}$ & $\approx0$ (DCSD) \\
{}~$D^0\to K^- K^+$ &
-0.11$<$$A$$<$0.16~\cite{Frabetti} & $10^{-3}$ & \\
& -0.028$<$$A$$<$0.166~\cite{Bartelt} & & \\
{}~$D^+\to K^- K^+\pi^+$ & -0.14$<$$A$$<$0.081~\cite{Frabetti}
& $10^{-3}$ & \\
{}~$D^+\to \overline {K^{*0}}K^+$
& -0.33$<$$A$$<$0.094~\cite{Frabetti} & $10^{-3}$ & $(2.8\!\pm\!0.8)\,\times\,
10^{-3}$~\cite{Pugliese} \\
{}~$D^+\to \phi\pi^+$ & -0.075$<$$A$$<$0.21~\cite{Frabetti}
& $10^{-3}$ & \\
{}~$D^+\to \eta\pi^+$ & & &
(-$1.5\!\pm\!0.4)\,\times\,10^{-3}$~\cite{Pugliese}\\
{}~$D^+\to K_S\pi^+$ & & few$\times10^{-4}$ & $3.3\times10^{-3}$~\cite{Xing} \\
\hline
FCNC
& & & \\ \hline
{}~$D^0\to\mu^+\mu^-$ & $7.6\times 10^{-6}$~\cite{WA92} &$10^{-7}$
& $<3\times10^{-15}$~\cite{Hewett} \\
{}~$D^0\to \pi^0\mu^+\mu^-$ & $1.8\times10^{-4}$~\cite{E653} & $10^{-6}$ & \\
{}~$D^0\to \overline {K^0} e^+e^-$ & $17.0\times10^{-4}~\cite{Sheldon}$
& $10^{-6}$ & $<2\times10^{-15}$~\cite{Hewett} \\
{}~$D^0\to\overline {K^0}\mu^+\mu^-$ & $2.6\times10^{-4}$~\cite{E653}
& $10^{-6}$ & $<2\times10^{-15}$~\cite{Hewett} \\
{}~$D^+\to \pi^+e^+e^-$ & $6.6\times10^{-5}$~\cite{Aitala}
& few$\,\times10^{-7}$ & $<10^{-8}$~\cite{Hewett} \\
{}~$D^+\to \pi^+\mu^+\mu^-$ & $1.8\times10^{-5}$~\cite{Aitala}
& few$\,\times10^{-7}$ & $<10^{-8}$~\cite{Hewett} \\
{}~$D^+\to K^+ e^+e^-$ & $4.8\times10^{-3}$~\cite{Sheldon}
& few$\,\times10^{-7}$ & $<10^{-15}$~\cite{Hewett} \\
{}~$D^+\to K^+ \mu^+\mu^-$ & $8.5\times10^{-5}$~\cite{PDG}
& few$\,\times10^{-7}$ & $<10^{-15}$~\cite{Hewett} \\
{}~$D\to X_u+\gamma$ & & & $\sim$10$^{-5}$~\cite{Hewett} \\
{}~$D^0\to \rho^0\gamma$ & $1.4\times10^{-4}$~\cite{Hewett} & &
$(1-5)\,\times\,10^{-6}$~\cite{Hewett} \\
{}~$D^0\to \phi\gamma$ & $2\times10^{-4}$~\cite{Hewett} &
& $(0.1-3.4)\,\times\,10^{-5}$~\cite{Hewett} \\
\hline
LF or LN Viol.\
& & & \\ \hline
{}~$D^0\to\mu^\pm e^\mp$ & $1.0\times 10^{-4}$~\cite{PDG} & $10^{-7}$ & 0 \\
{}~$D^+\to\pi^+\mu^\pm e^\mp$ & $3.2\times 10^{-3}$~\cite{Sheldon}
& few$\times10^{-7}$ & 0 \\
{}~$D^+\to K^+ \mu^\pm e^\mp$ & $3.3\times 10^{-3}$~\cite{Sheldon}
& few$\times10^{-7}$ & 0 \\
{}~$D^+\to \pi^- \mu^+\mu^+$ & $2.2\times 10^{-4}$~\cite{E653}
& few$\times10^{-7}$ & 0 \\
{}~$D^+\to K^- \mu^+\mu^+$ & $3.4\times 10^{-4}$~\cite{E653}
& few$\times10^{-7}$ & 0 \\
{}~$D^+\to \rho^- \mu^+\mu^+$ & $5.6\times 10^{-4}$~\cite{E653}
& few$\times10^{-7}$ & 0 \\
\hline
Mixing
& & & \\ \hline
{}~${}^{^{(}}{\overline {D^0}}{}^{^{)}}\to K^\mp\pi^\pm$ &
$r<0.37$\%~\cite{E691}, & $r<10^{-5},$ & \\
& $|\Delta M_D|<1.3\!\times\!10^{-4}$\,eV & $|\Delta M_D|<10^{-5}\,$eV &
$10^{-7}$\,eV~\cite{Burdman} \\
{}~${}^{^{(}}{\overline {D^0}}{}^{^{)}}\to \ell\nu X$ & $r <
0.56$\%~\cite{Louis}
 & & \\
{}~${}^{^{(}}{\overline {D^0}}{}^{^{)}}\to K\ell\nu$ & & $r<10^{-5}$ & \\
\hline
\end{tabular}
\end{center}
$^*$at 90\% confidence level
\end{table}
}

\begin{table}\centering
\footnotesize
\caption
{ Estimated yields of reconstructed events (antiparticles
included)}
\label{tab:yields}
a) direct estimates
\begin{tabular}{|l|c|c|c|c|c|}
\hline
mode & charm frac. &
BR (\%) & acceptance & efficiency & yield \\
\hline
\hline
$D^0\to K^-\pi^+$ & 0.5 & 4.0 & 0.6 & 0.1 & $1.3\times10^8$ \\
$D^+\to K^{*0}\mu\nu$ & 0.25 & 2.7 & 0.4 & 0.25 & $7\times10^7$\\
\qquad$\to K\pi\mu\nu$ &&&&&\\
all & 1 & $\approx0.1$ & $\approx 0.4$ & $\approx 0.1$ &
$\approx4\times10^8$\\
\hline
\end{tabular}
\\
b) extrapolations from E791
\\
\begin{tabular}{|l|c|c|c|c|}
\hline
mode & BR (\%) & E791 yield & Charm2000 yield & analysis\\
\hline
\hline
$D^+\to K^-\pi^+\pi^+$ & 9.1 & $37000\pm200$ & $(7\pm0.001)\times10^7$ & FCNC
\\
$D^+\to K_S\pi^+$ & 0.94 & & $(7\pm0.003)\times10^6$ & \\
$D^{*+}\to \pi^+D^0\to\pi^+K^-\pi^+$ & 2.7 & 5000 & $10^7$ & mixing \\
$D^{*+}\to \pi^+D^0\to\pi^+K^-\pi^+\pi^+\pi^-$ & 5.5 & 3200 & $0.6\times10^7$
& mixing \\
$D^{*+}\to \pi^+D^0\to\pi^+K^+\pi^-$ & 0.02? & 45? & $10^4-10^5$ & DCSD \\
$D^0\to K^-\pi^+\pi^+\pi^-$ & 8.1 & & $6\times10^7$ & \\
\hline
\end{tabular}
\end{table}

\end{document}